\documentclass[12pt,onecolumn,prc,showpacs,preprintnumber,amsmath,
floatfix,amssymb]{revtex4}
\usepackage{epsfig}
\usepackage{graphicx}
\usepackage{dcolumn}
\usepackage{bm}
\def\r{\mbox{{\bf  r}}}
\def\p{\mbox{\boldmath $p$}}
\def\q{\mbox{\boldmath $q$}}
\def\k{\mbox{\boldmath $k$}}
\def\t{\mbox{\boldmath $t$}}
\allowdisplaybreaks[4]

\begin{document}
\title{Quasi-elastic neutrino charged-current scattering off $^{12}$C}
\author{A.~V.~Butkevich}
\affiliation{ Institute for Nuclear Research,
Russian Academy of Sciences,
60th October Anniversary Prosp. 7A,
Moscow 117312, Russia}
\date{\today}
\begin{abstract}

The charged-current quasi-elastic scattering of muon neutrino on a carbon
target is calculated for neutrino energy up to 2.8 GeV using
the relativistic distorted-wave impulse approximation with relativistic
optical potential, which was earlier successfully applied to describe 
electron-nucleus data. We studied both neutrino and electron processes and 
have shown that the reduced exclusive cross section for neutrino and electron 
scattering are similar. 
We have also studied nuclear and axial vector mass effects on the shape 
of $Q^2$ distribution. The comparison of the (anti)neutrino total cross
sections per (proton)neutron, calculated for the carbon and oxygen targets 
shows the cross sections for oxygen to be lower than those for carbon. We 
found significant nuclear model dependence of inclusive and total cross 
sections for energy about 1 GeV.
\end{abstract}

\maketitle

\section{Introduction}

The goals of the current and planed set of accelerator-based neutrino 
experiments~\cite{MiniA,Sci,Mino,Miner,T2K,OPERA,NOVA} are the precision 
measurements of the neutrino mass squared difference $\Delta m^2_{23}$ by 
measuring muon neutrino disappearance, and searching for the last unmeasured 
leptonic mixing angle $\theta_{13}$ through the muon to electron neutrino 
transition. The last oscillation channel, if it occurs, opens the possibility 
of observation matter/anti-matter asymmetries in neutrinos and
determination of the ordering of the neutrino mass states. 
The data of these experiments will greatly extend the statistics due to
extremely intense neutrino beamline.

To study the neutrino oscillation effects on the terrestrial distance
scale, the neutrino beams cover the energy range from a
few hundred MeV to several GeV. In this energy range, the dominant 
contribution to the neutrino-nucleus cross section comes from the 
charged-current (CC) quasi-elastic (QE) reactions and resonance 
production processes. The cross section data in this energy range are rather 
scarce and were taken on the targets, which are not used in the neutrino 
oscillation experiments (i.e.  water, iron, lead or plastic).
In this situation, the statistical uncertainties should be negligible as 
compared to systematic errors in the incident neutrino flux, neutrino 
interaction model and the detector effects on the neutrino events selection and
neutrino energy reconstruction. Apparently, these uncertainties produce
systematic errors in the extraction of oscillation parameters.

Many experiments try to reduce these uncertainties by using a near detectors. 
One of the option for near detector design is to make the near detector as 
more segmented and fine-grained, using scintillator (carbon) as a target and 
detecting material. This strategy means that one must try to measure the
fluxes and cross sections as independently as possible, and then use this
information to constrain the detector simulation so that the information is
correctly extrapolated to the far detector. The concern with this strategy is
that the detector simulation must accurately predicted the detector
response. Because the near and far detectors are not
necessarily of the same target material, a part of the near detector must
include some of the same target material, so that nuclear effects on the cross
sections (QE and non-QE) could be taking into account. Among the proposed 
experiments MINERvA~\cite{Miner} and ND280 detector~\cite{ND} will have 
the possibility for precise measurements in a wide range of energies and for  
various nuclear targets.     

To model the QE neutrino scattering from a nuclei, 
the most part of the Monte Carlo (MC) event generators~\cite{Zel} are based 
on the relativistic Fermi gas model (RFGM)~\cite{Moniz} with Pauli blocking, 
in which the nucleus is described as a system of quasi-free nucleons with a 
flat nucleon momentum distribution up to the same Fermi momentum $p_F$ and 
nuclear binding energy $\epsilon_b$. But this model does not take into account 
the nuclear shall structure, the final state interaction (FSI) between the 
outgoing nucleon and residual nucleus and the presence of short-range 
nucleon-nucleon $(NN)$ correlations, leading to appearance of a high-momentum 
and high-energy component in the nucleon momentum-energy distribution in the 
target.

The comparison with the high-precision electron scattering data 
has shown ~\cite{But1} that the accuracy of the RFGM prediction becomes 
poor at low squared four-momentum transfer $Q^2$, where the nuclear effects 
are largest. The modern quasi-elastic neutrino scattering data (the CC QE 
event distribution as a function of $Q^2$)~\cite{MiniA,Gran} also reveal the 
inadequacies in the present neutrino cross section simulation. The 
data/MC disagreement shows the data deficit in the low-$Q^2$ ($Q^2 \leq 0.2$ 
(GeV/c)$^2$) region.

There are many calculations for the QE neutrino charged-current and
neutral-current scattering cross sections on nucleus, which go beyond the
simple RFGM and use more realistic description of nuclear dynamics. In
calculation of Refs.\cite{Benh1,Benh2} within the plane-wave impulse 
approximation (PWIA), the short range $NN$-correlations were included using 
the description of nuclear dynamics, based on nuclear many body theory. 
Charged current and/or neutral current neutrino-nucleus cross sections were 
studied within the relativistic distorted-wave impulse approximation (RDWIA) 
in Refs.\cite{Meu1,Meu2,Meu3,Mair,Ryck1,Ryck2,But2,But3,Kim1,Kim2}, 
using the relativistic shell model approach and taking into account the FSI 
effects. In Refs.\cite{But2,But3} the contribution of the short range 
correlations (SRC) was also considered. The FSI effects were studied in 
Refs.\cite{Ni1,Ni2,Sin1,Sin2} within the framework of the random phase 
approximation, in Refs.\cite{Don1,Don2,Don3} - within a Superscaling approach, 
and in Ref.\cite{Lei} - in a GiBUU model. 

In this paper, we calculate the single-nucleon knockout contribution to the
exclusive, inclusive and total cross sections of charged-current QE
(anti)neutrino scattering from ${}^{12}$C, using different approximations (PWIA
and RDWIA) and the Fermi gas model. We employ the LEA code~\cite{LEA} which was
adopted for neutrino reactions. In our approach, the effect of the SCR in
the carbon ground state is evaluated in the PWIA~\cite{Ciofi,Kul} and the FSI 
effect on the inclusive cross sections in the presence of the 
$NN$-correlations is estimated according Ref.\cite{But2}.  
The aims of this work are a) calculation the RDWIA CC QE 
$\nu {}^{12}$C cross sections, b) investigation of nuclear effects on 
the $Q^2$ dependence of the (anti)neutrino cross section, and 
c) comparison of the total cross sections, scaled with the number of 
neutron/proton in the target for (anti)neutrino scattering on the oxygen and
carbon targets.

The outline of this paper is the following. In Sec. II we present briefly the
formalism for the CC QE scattering process and the RDWIA model. The results 
are presented and discussed in Sec. III. Our conclusions are summarized in 
Sec. IV. 

\section{Formalism of quasi-elastic scattering and RDWIA}

We consider electron and neutrino charged-current QE exclusive
\begin{equation}\label{qe:excl}
l(k_i) + A(p_A)  \rightarrow l^{\prime}(k_f) + N(p_x) + B(p_B),      
\end{equation}
and inclusive
\begin{equation}\label{qe:incl}
l(k_i) + A(p_A)  \rightarrow l^{\prime}(k_f) + X                      
\end{equation}
scattering off nuclei in one-photon (W-boson) exchange approximation. Here $l$
labels the incident lepton [electron or muon (anti)neutrino], and
$l^{\prime}$ represents the scattered lepton (electron or muon),
$k_i=(\varepsilon_i,\k_i)$ 
and $k_f=(\varepsilon_f,\k_f)$ are the initial and final lepton 
momenta, $p_A=(\varepsilon_A,\p_A)$, and $p_B=(\varepsilon_B,\p_B)$ are 
the initial and final target momenta, $p_x=(\varepsilon_x,\p_x)$ is the 
ejectile nucleon momentum, $q=(\omega,\q)$ is the momentum transfer carried by 
the virtual photon (W-boson), and $Q^2=-q^2=\q^2-\omega^2$ is the photon 
(W-boson) virtuality. 

\subsection{ CC QE neutrino-nucleus cross sections}

In the laboratory frame, the differential cross section for the exclusive
electron ($\sigma ^{el}$) and (anti)neutrino CCQE ($\sigma ^{cc}$) scattering, 
in which only a single discrete state or narrow resonance of the target is 
excited, can be written as
\begin{subequations}
\label{cs5}
\begin{align}
\frac{d^5\sigma^{el}}{d\varepsilon_f d\Omega_f d\Omega_x} &= R
\frac{\vert\p_x\vert{\varepsilon}_x}{(2\pi)^3}\frac{\varepsilon_f}
{\varepsilon_i} \frac{\alpha^2}{Q^4} L_{\mu \nu}^{(el)}W^{\mu \nu (el)}
\\                                                                       
\frac{d^5\sigma^{cc}}{d\varepsilon_f d\Omega_f d\Omega_x} &= R
\frac{\vert\p_x\vert{\varepsilon}_x}{(2\pi)^5}\frac{\vert\k_f\vert}
{\varepsilon_i} \frac{G^2\cos^2\theta_c}{2} L_{\mu \nu}^{(cc)}W^{\mu \nu (cc)},
\end{align}
\end{subequations}
 where $\Omega_f$ is the solid angle for the lepton momentum, $\Omega_x$ is the
 solid angle for the ejectile nucleon momentum, $\alpha\simeq 1/137$ is the 
fine-structure constant, $G \simeq$ 1.16639 $\times 10^{-11}$ MeV$^{-2}$ is
the Fermi constant, $\theta_C$ is the Cabbibo angle
($\cos \theta_C \approx$ 0.9749), $L_{\mu \nu}$ is the lepton tensor and 
$W^{(el)}_{\mu \nu}$ and $W^{(cc)}_{\mu \nu}$ are,
respectively, the electromagnetic and weak CC nuclear tensors. The recoil 
factor $R$ is given by
\begin{equation}\label{Rec}
R =\int d\varepsilon_x \delta(\varepsilon_x + \varepsilon_B - \omega -m_A)=
{\bigg\vert 1- \frac{\varepsilon_x}{\varepsilon_B}
\frac{\p_x\cdot \p_B}{\p_x\cdot \p_x}\bigg\vert}^{-1},                    
\end{equation}
 and $\varepsilon_x$ is the solution to the equation
\begin{equation}\label{eps}
\varepsilon_x+\varepsilon_B-m_A-\omega=0,                                 
\end{equation}
where $\varepsilon_B=\sqrt{m^2_B+\p^2_B}$, $~\p_B=\q-\p_x$, $~\p_x=
\sqrt{\varepsilon^2_x-m^2}$, and $m_A$, $m_B$, and $m$ are masses of the 
target, recoil nucleus and nucleon, respectively. 
The missing momentum $p_m$ and missing energy $\varepsilon_m$ are defined by 
\begin{subequations}
\begin{align}
\label{p_m}
\p_m & = \p_x-\q
\\
\label{eps_m}
\varepsilon_m & = m + m_B - m_A                                           
\end{align}
\end{subequations}
The leptonic tensor is separated into a symmetrical and an anti-symmetrical
components that are written as in Ref.~\cite{But2}. 
The electromagnetic and weak CC hadronic tensors, $W^{(el)}_{\mu \nu}$ and 
$W^{(cc)}_{\mu \nu}$ are given by bilinear products 
of the transition matrix elements of the nuclear electromagnetic or CC 
operator $J^{(el)(cc)}_{\mu}$ between the initial nucleus state 
$\vert A \rangle $ and the final state $\vert B_f \rangle$ as
\begin{eqnarray}
W^{(el)(cc)}_{\mu \nu } &=& \sum_f \langle B_f,p_x\vert          
J^{(el)(cc)}_{\mu}\vert A\rangle \langle A\vert
J^{(el)(cc)\dagger}_{\nu}\vert B_f,p_x\rangle,              
\label{W}
\end{eqnarray}
where the sum is taken over undetected states.

In the inclusive reactions (\ref{qe:incl}) only the outgoing lepton is
detected, and the differential cross sections can be written as
\begin{subequations}\label{csinc}
\begin{align}
\frac{d^3\sigma^{el}}{d\varepsilon_f d\Omega_f} &=
\frac{\varepsilon_f}{\varepsilon_i}
 \frac{\alpha^2}{Q^4} L_{\mu \nu}^{(el)}\mathcal{W}^{\mu \nu (el)},
\\                                                                       
\frac{d^3\sigma^{cc}}{d\varepsilon_f d\Omega_f } &=
\frac{1}{(2\pi)^2}\frac{\vert\k_f\vert}
{\varepsilon_i} \frac{G^2\cos^2\theta_c}{2} L_{\mu \nu}^{(cc)}
\mathcal{W}^{\mu \nu (cc)},
\end{align}
\end{subequations}
where $\mathcal{W}^{\mu \nu}$ is the inclusive hadronic tensor.
The expressions for the exclusive (\ref{cs5}) and inclusive
(\ref{csinc}) lepton scattering cross sections in terms of response
functions are given in Ref.\cite{But2}.

It is also useful to define a reduced cross section
\begin{equation}
\sigma_{red} = \frac{d^5\sigma}{d\varepsilon_f d\Omega_f d\Omega_x}
/K^{(el)(cc)}\sigma_{lN},                                                             
\end{equation}
where
are phase-space factors for electron and neutrino scattering, the recoil
factor $R$ is given by Eq.(\ref{Rec}), and $\sigma_{lN}$ is the corresponding
elementary cross section for the lepton scattering from the moving free
nucleon.

\subsection{ Models}

We describe the lepton-nucleon scattering in the impulse 
approximation (IA), in which only one nucleon of the target is involved in 
the reaction, and  the nuclear current is written as a sum of single-nucleon 
currents. Then, the nuclear matrix element in Eq.(\ref{W}) takes the form
\begin{eqnarray}\label{Eq12}
\langle p,B\vert J^{\mu}\vert A\rangle &=& \int d^3r~ \exp(i\t\cdot\r)
\overline{\Psi}^{(-)}(\p,\r)
\Gamma^{\mu}\Phi(\r),                                                     
\end{eqnarray}
where $\Gamma^{\mu}$ is the vertex function, $\t=\varepsilon_B\q/W$ is the
recoil-corrected momentum transfer, $W=\sqrt{(m_A + \omega)^2 - \q^2}$ is the
invariant mass, $\Phi$ and $\Psi^{(-)}$ are relativistic bound-state and
outgoing wave functions.

For electron scattering, we use the CC2 electromagnetic vertex
function for a free nucleon~\cite{deFor}
\begin{equation}
\Gamma^{\mu} = F^{(el)}_V(Q^2)\gamma^{\mu} + {i}\sigma^{\mu \nu}\frac{q_{\nu}}
{2m}F^{(el)}_M(Q^2),                                                      
\end{equation}
where $\sigma^{\mu \nu}=i[\gamma^{\mu},\gamma^{\nu}]/2$, $F^{(el)}_V$ and
$F^{(el)}_M$ are the Dirac and Pauli nucleon form factors. 
The single-nucleon charged current has $V{-}A$ structure $J^{\mu(cc)} = 
J^{\mu}_V + J^{\mu}_A$. For the free-nucleon vertex function 
$\Gamma^{\mu(cc)} = \Gamma^{\mu}_V + \Gamma^{\mu}_A$ we use the CC2 vector 
current vertex function
\begin{equation}
\Gamma^{\mu}_V = F_V(Q^2)\gamma^{\mu} + {i}\sigma^{\mu \nu}\frac{q_{\nu}}
{2m}F_M(Q^2)                                                            
\end{equation}
and the axial current vertex function
\begin{equation}
\Gamma^{\mu}_A = F_A(Q^2)\gamma^{\mu}\gamma_5 + F_P(Q^2)q^{\mu}\gamma_5.  
\end{equation}
The weak vector form factors $F_V$ and $F_M$ can be expressed in terms of
the corresponding electromagnetic factors for proton $F^{(el)}_{i,p}$ and 
neutron $F^{(el)}_{i,n}$ as follows
\begin{equation}
F_i = F^{(el)}_{i,p} - F^{(el)}_{i,n}.                                 
\end{equation}
For the electromagnetic and weak CC vector vertexes we employ the de Forest
prescription~\cite{deFor} (because the bound nucleons are off shell) and
Coulomb gauge. 
For the Dirac and Pauli nucleon form factors we use the 
approximation from Ref.~\cite{MMD} and the dipole approximation for the axial 
$F_A$ and psevdoscalar $F_P$ form factors 
\begin{equation}
F_A(Q^2)=\frac{F_A(0)}{(1+Q^2/M_A^2)^2},\quad                          
F_P(Q^2)=\frac{2m F_A(Q^2)}{m_{\pi}^2+Q^2},
\end{equation}
where $F_A(0)=1.267$ and $m_{\pi}$, $M_A$ are the pion and axial mass, 
respectively. 

In Ref.~\cite{Kel}, a formalism was developed for the 
$A({e},e^{\prime}N)B$ reaction
that describes channel coupling in the FSI of the $N+B$ system. In this
work the independent particle shell model (IPSM) is assumed for nuclear
structure. The model space for
$^{12}$C$(l,l^{\prime}N)$ consists of $1s_{1/2}$ and $1p_{3/2}$
nucleon-hole states in the $^{11}$B and $^{11}$C nuclei.
The $1s_{1/2}$ state is regarded as a discrete state even though its spreading
width is actually appreciable. 

In the independent particle shell model the relativistic bound-state functions 
$\Phi$ in Eq.(\ref{Eq12}) are obtained within the Hartree--Bogolioubov 
approximation in the $\sigma-\omega$ model ~\cite{Serot}.  
The upper component of the bound-state wave function $\Phi$ is used for 
calculation of the shell nucleons spectral function in the PWIA calculations.
We use the nucleon 
bound-state functions calculated by the TIMORA code~\cite{HS} with the
normalization factors $S(\alpha)$ relative to full occupancy of the IPSM 
orbitals of ${}^{12}$C: $S(1p_{3/2})$=84\%, 
$S(1s_{1/2})$=100\%, and an average factor of about 89\%. These estimations of
the depletion of hole states follow from the RDWIA analysis of 
${}^{12}$C$({e},e^{\prime}{p})$ for $Q^2 < 2$ (GeV/c)$^2$~\cite{Kel2}
and consist with a direct measurement of the spectral function using 
${}^{12}$C$({e},e^{\prime}{p})$ in parallel kinematics~\cite{Rohe}, which 
observed approximately 0.6 protons in a region with $p_m \geq$ 240 Mev/c and 
$\varepsilon_m \geq$ 50 MeV attributable to a single-nucleon knockout from
correlated cluster. Similar estimates of the depletion of hole states are
available from the self-consistent Green's function method~\cite{Frick},
correlated basis function theory~\cite{Benh3} and other method also. 
%

In the RDWIA the ejectile wave function $\Psi$ in Eq.(\ref{Eq12}) is obtained 
following the direct Pauli reduction method~\cite{Udi,Hed}. It is well known 
that the Dirac spinor
\begin{eqnarray}
\Psi = 
\begin{pmatrix}                                                          
\Psi _+ \\ 
\Psi_-
\end{pmatrix}
\end{eqnarray}
can be written in terms of its positive energy component $\Psi _+$ as
\begin{eqnarray}
\Psi = 
\begin{pmatrix}  \Psi _+ \\
 \frac {\bm{\sigma} \cdot \p}{E+M+S-V} \Psi _+                          
\end{pmatrix},
\end{eqnarray}
where $S=S(r)$ and $V=V(r)$ are the scalar and vector potentials for the
 nucleon with energy $E$. The upper component $\Psi _+$ can be related to the  
Schr\"odinger-like wave function $\xi $ by the Darwin factor $D(r)$, i.e.
\begin{eqnarray}
\Psi _+ = \sqrt {D(r)}\ \xi  \label{eq.scheq},
\end{eqnarray}                                                            
\begin{eqnarray}
D(r) = \frac {E+M+S(r)-V(r)}{E+M} \label{eq.Darwin}.                      
\end{eqnarray}
The two-component wave function $\xi $ is the solution of the Schr\"odinger 
equation containing equivalent central and spin-orbit potentials, which are 
functions of the scalar and vector potentials $S$ and $V$, and are energy 
dependent. We use the LEA program~\cite{LEA} for numerical calculation of the
distorted wave functions with EDAD1 SV relativistic optical potential
\cite{Coop} for carbon.

In the Plane-Wave Impulse Approximation (PWIA) the final state interaction 
between the outgoing nucleon and the residual nucleus is neglected, and the 
nonrelativistic PWIA exclusive cross section has a factorized form~\cite{Frul}
\begin{equation}
\frac{d^5\sigma}{d \varepsilon_f d\Omega_f d\Omega_x} = 
K^{(el)(cc)}\sigma_{lN}\mathcal{P}(E,\p) 
\end{equation}
where $\mathcal{P}(E,\p)$ is the nuclear spectral function.   

According to the JLab data \cite{Dutta,Rohe}, the occupancy of the 
independent particle shell model orbitals of $^{12}$C equals about 89\%, on 
the average. In this work we assume that the missing strength (11\%) can be 
attributed to the short-range $NN$-correlations in the ground state, leading to
appearance of high-momentum (HM) and high-energy nucleon distribution in the 
target. In order to estimate this effect in the inclusive cross sections, 
we consider the phenomenological model~\cite{Ciofi,Kul}
 where the high-momentum (HM) part of the spectral function is
determined by excited states with one or more nuclei in a continuum.

We calculate the inclusive cross sections with the FSI effects in the presence
of the short-range $NN$-correlations, using the approach which was proposed in
Ref.\cite{But2}. 

\section{Results and analysis}

\subsection{Electron scattering}

The LEA code was successfully tested against ${}^{12}$C$(e,e^{\prime}p)$ data. 
For illustration, Fig.1 shows measured JLab~\cite{Dutta} reduced cross 
sections for the removal of protons from the $1s$ and $1p$ shells of
$^{12}$C as functions of missing momentum $p_m$ as compared with LEA code
calculations. 
\begin{figure*}
  \begin{center}
    \includegraphics[height=10cm,width=18cm]{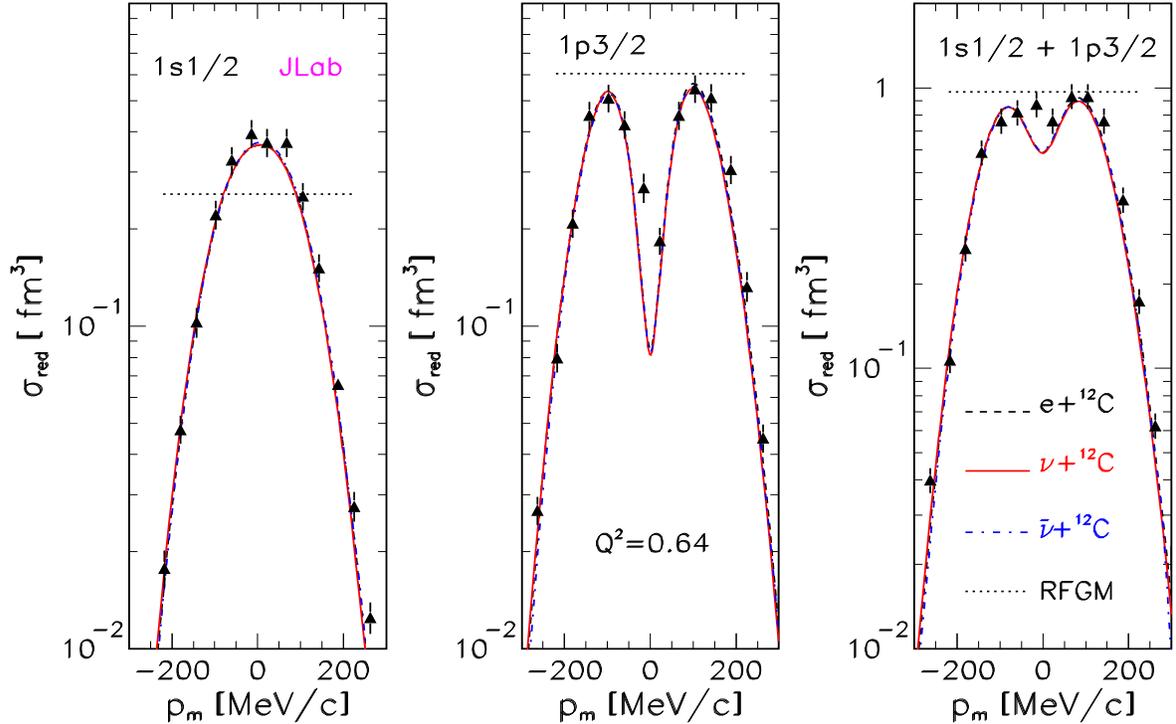}
  \end{center}
  \caption{(Color online) Comparison of the RDWIA and the RFGM calculations 
for electron, neutrino and antineutrino reduced  cross sections for the 
removal of nucleons from 1p and 1s shells of $^{12}$C as functions of missing 
momentum. JLab data~\cite{Dutta} for beam energy $E_{beam}$=2.455 GeV, proton
kinetic energy $T_p$=350 MeV, and $Q^2$=0.64 (GeV/c)$^2$. The RDWIA
calculations are shown for electron scattering (dashed line) and neutrino
(solid line) and antineutrino (dashed-dotted line) scattering; and the RFGM 
results are shown for the reduced cross sections (dotted line) for the JLab
kinematics.} 
\end{figure*}
It should be noted that
negative values of $p_m$ correspond to $\phi=\pi$ and positive ones to
$\phi$=0, where $\phi$ is the angle between the scattering ($\k_i,\k_f$) and 
reaction ($\p_x,\p_B$) planes. The data for beam energy $E_{beam}$=2.445 GeV 
and $Q^2$=0.6, 1.2, and 1.8 (GeV/c)$^2$ were measured in the 
quasi-perpendicular kinematics with constant ($\omega,\q$). The detailed 
analysis data~\cite{Kel2} for ${}^{12}$C$(e,e^{\prime}p)$ with 
$Q^2\leq$2 (GeV/c)$^2$ using the RDWIA based upon Dirac-Hartree wave 
functions has shown that the 1p normalization extracted from data for 
$Q^2\geq$0.6 (GeV/c)$^2$ is equals approximately 0.87, independent of $Q^2$. 
The total 1p and 1s strength for $\varepsilon_m \leq $80 MeV approaches 100\% 
of IPSM, consistent with a continuum contribution for 
30$\leq \varepsilon_m\leq$80 MeV of about 12\% of IPSM.

The electron and neutrino scattering off the nuclei are closely interrelated 
and one can treat both processes within the same formalism. In the
nonrelativistic PWIA, $\sigma_{red}$ is a nuclear spectral function and 
should be similar for electron and (anti)neutrino scattering except small 
distinctions which can be attributed to the Coulomb distortion upon the 
electron wave function. The small difference between neutrino and antineutrino
is due to difference in the FSI of the proton and neutron with the residual
nucleus. This effect is neglected at the energy beam higher than 1 GeV. 
There is an overall good agreement between calculated in the RDWIA electron 
and (anti)neutrino cross sections and data. Apparently the RFGM predictions 
(with the Fermi momentum $p_F$=221 MeV/c and binding energy 
$\epsilon_b$=25 MeV) overestimate the values of cross sections and completely 
off the exclusive data. This is due to the uniform momentum distribution of 
the Fermi gas model and neglecting by the FSI effects. Therefore,
the RFGM can not predict well enough the momentum distribution of outgoing
protons in simulation of the CC QE two-track events at momentum transfer 
$|\q|\leq |\p_m|$, i.e. at low $Q^2$. 
\begin{figure*}
  \begin{center}
    \includegraphics[height=16cm,width=16cm]{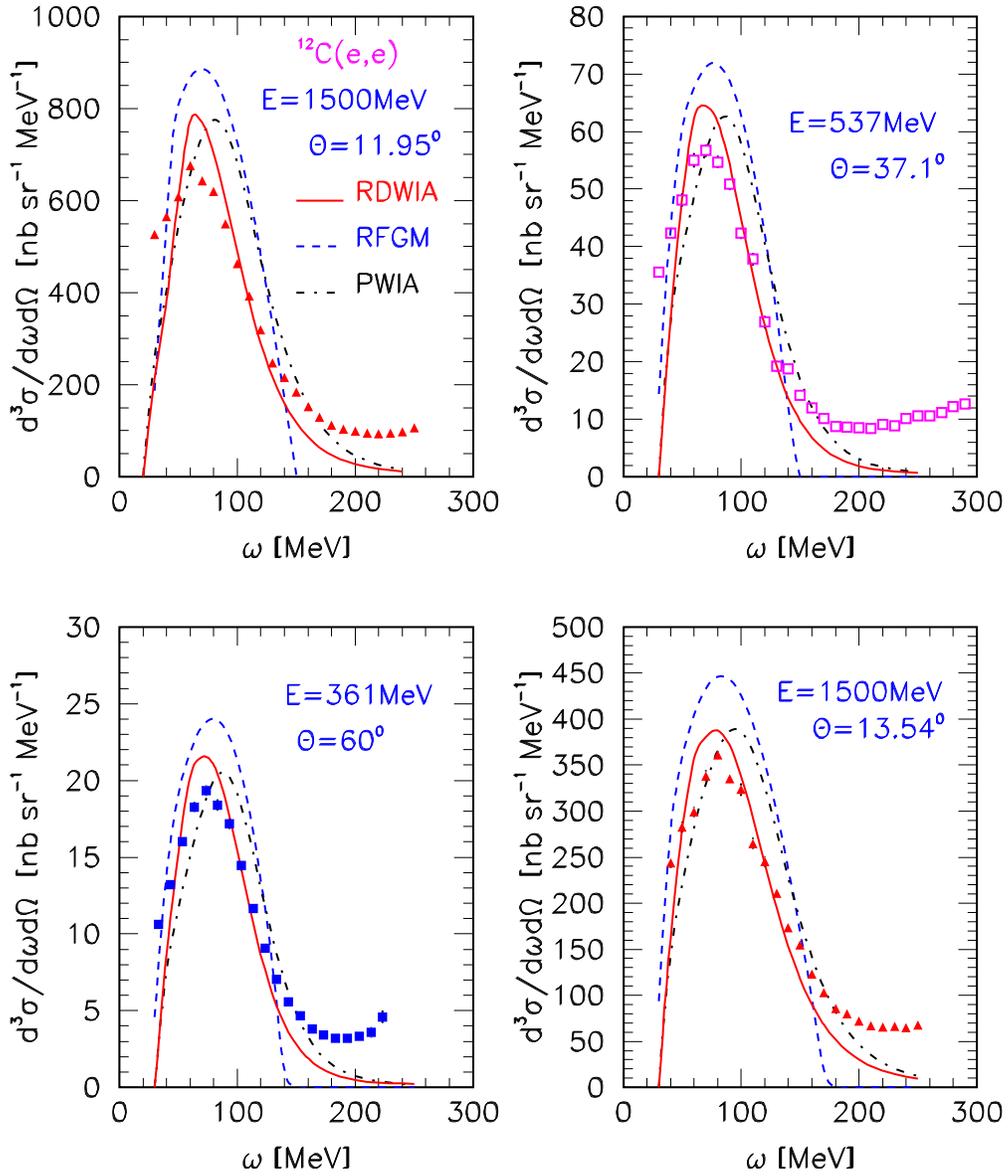}
  \end{center}
  \caption{(Color online) Inclusive cross section versus energy transfer
$\omega$ for electron scattering on $^{12}$C. The data are from 
Ref.\cite{Baran} (filled triangles), Ref.\cite{Conn} (open squares), 
and Ref.\cite{Barr} (filled squares). In Ref.\cite{Baran} data are for the 
electron beam energy $E_e$=1500 MeV, and scattering angles
$\theta_e$=11.95$^{\circ}$, 13.54$^{\circ}$; in Ref.\cite{Conn} data are for 
$E_e$=537 MeV and $\theta_e$=37.1$^{\circ}$; in Ref.\cite{Barr} data are for 
$E_e$=361 MeV and $\theta_e$=60$^{\circ}$.}
\end{figure*}
\begin{figure}
  \begin{center}
    \includegraphics[height=16cm,width=16cm]{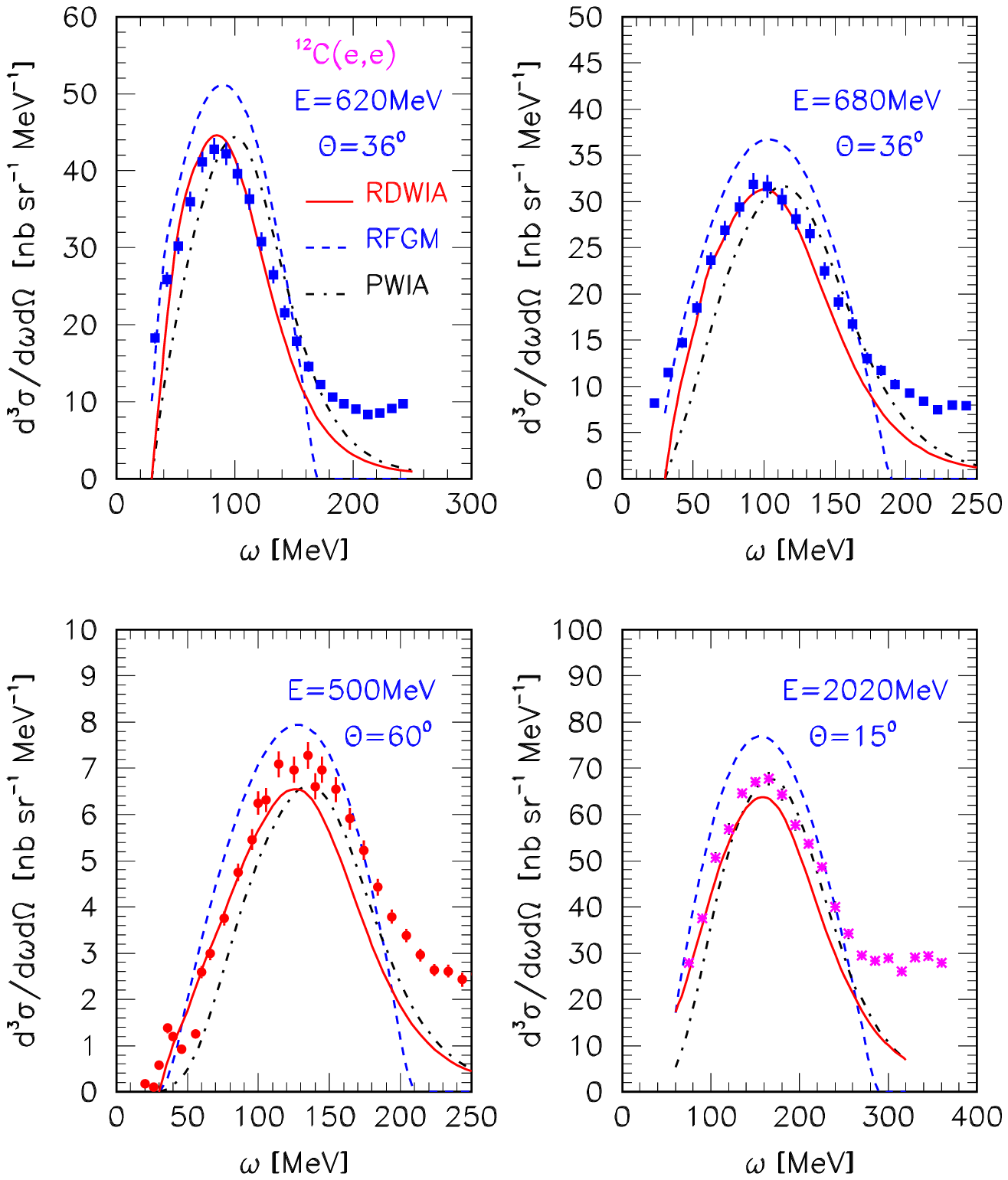}
  \end{center}
  \caption{(Color online) Same as Fig.2, but the data are from Ref.\cite{Barr}
    (filled squares) for $E_e$=620 MeV, $\theta_e$=36$^{\circ}$ and 
$E_e$=680 MeV, $\theta_e$=36$^{\circ}$; Ref.\cite{Whit} (filled circles) for
$E_e$=500 MeV, $\theta_e$=60$^{\circ}$; Ref.\cite{Day} (stars) for 
$E_e$=2020 MeV, $\theta_e$=15$^{\circ}$.}   
\end{figure}

A complex relativistic optical potential with a nonzero imaginary part
generally produces an absorption of flux. However, for the inclusive reaction, 
the total flux must conserve. Currently there is no fully consistent solution 
to this problem, and different approaches are used. In Refs.\cite{Meu1,Meu4} 
it was shown that the inclusive CC cross sections calculated with only the 
real part of the optical potential are almost identical those of the 
Green's function approach, in which the FSI effect in the inclusive
reactions is treated by means of a complex optical potential and the total
flux is conserved. In this work, in order to calculate the inclusive and total 
cross sections, we use the approach, in which only the real part of the 
optical potential EDAD1 is included. 

To test our approach, we calculated the inclusive
$^{12}$C$(e,e^{\prime})$ cross sections and compared them with data from SLAC
~\cite{Baran,Conn,Whit}, from Saclay~\cite{Barr}, and from 
JLab~\cite{Day}. Figures 2 and 3 show measured inclusive cross sections 
as functions of energy transfer as compared to the RDWIA, PWIA, and RFGM 
calculations. These data cover the range of the tree-momentum transfer 
(around the peak) form $|\q|\approx$310 MeV/c (beam energy $E_e$=1500 MeV and 
scattering angle $\theta$=11.95$^\circ$) up to $|\q|\approx$530 MeV/c 
($E_e$=2020 MeV, $\theta$=15$^\circ$). We note that, relative to the PWIA 
results, the generic effect of the FSI with the real part of the optical 
potential is to reduce the cross section value around the peak and to shift 
the peak toward the lower value energy transfer. The inclusion of the 
high-momentum component increases the inclusive cross section in the 
high-energy transfer region and improves the agreement with data. For the 
RDWIA results, the difference between the calculated and measured cross 
sections at the maximum is less than $\pm$12\%. For the RFGM results these 
difference decreases with $\vert\q\vert$ from about 20\% at 
$\vert\q\vert\approx$310 MeV/c down to 
$\approx$13\% at $\vert\q\vert\approx$510 MeV/c. These results demonstrate
a strong  nuclear-model dependence of the inclusive cross sections at low
momentum transfer.
 
\subsection{Neutrino scattering}

The charged-current QE events distributions as functions of $Q^2$ were
measured in K2K~\cite{Gran, Espinal} and MiniBooNE~\cite{MiniA}
experiments. The shape of the $Q^2$ distribution, which is weakly dependents on
the flux uncertainties, was analyzed. High statistic data show a disagreement
with the RFGM predictions. The data samples exhibit significant deficit in the
region of low $Q^2\leq$0.2 (GeV/c)$^2$ (so call low-$Q^2$ problem). In
Ref.\cite{MiniA} it was shown that the data/MC disagreement is not due to
mis-modeling of the incoming neutrino energy spectrum, but to inaccuracy in
the simulation of CC QE interactions. To tune the Fermi gas model to the low
$Q^2$, an additional parameter $\kappa$ was introduced which reduced the phase
space volume of the nucleon Fermi gas at low-momentum transfer. This parameter
controls the $Q^2$ distribution in the low $Q^2$ region only.

In the region of high $Q^2$ the data excess is observed, and the value of the
axial vector mass $M_A$ obtained from a fit to the measured data, are higher 
than the results of previous experiments. The formal averaging of $M_A$ values 
from several experiments, which are very wide spread from 0.7 to 1.3 GeV, was 
done in Ref.\cite{Bern}: $M_A$=1.026$\pm$0.021. This result is also known as 
the axial mass world average value. K2K obtained the value of 1.2$\pm$0.12 
from the SciFi detector~\cite{Gran} using the water-aluminum mixture as a 
target, and also the preliminary result 1.14$\pm$0.11 from the SciBar detector
\cite{Espinal} using a scintillator target. The MiniBooNE experiment
(scintillator target) found that the data were better described with an
adjustment of two parameters $M_A$=1.23$\pm$0.20 GeV and
$\kappa$=1.019$\pm$0.011~\cite{MiniA}.

Recently the NOMAD experiment~\cite{Lyub} extracted the value of
$M_A$=1.05$\pm$0.02$\pm$0.06 GeV using a carbon target, which is in agreement
with the world average value. This result was obtained from the analysis of a
measured $\nu {}^{12}C$ total CC QE cross section for neutrino energy above 
$\approx$4 GeV, where the cross section 'plateaus' is reasonably well-known. 
It should be noted that both approaches, i.e. analysis of the shape of the 
$Q^2$ distribution and the direct measurement of the total cross section 
assume, that the vector form factors are known well from the electron 
scattering experiments. Actually, at $Q^2 \geq $ 3 (GeV/c)$^2$ 
the values of the neutron form factors are much less known that those of the 
proton~\cite{Bodek}, 
and the relative contribution from this region to the total cross section 
increases with neutrino energy.    
\begin{figure*}
  \begin{center}
    \includegraphics[height=16cm,width=16cm]{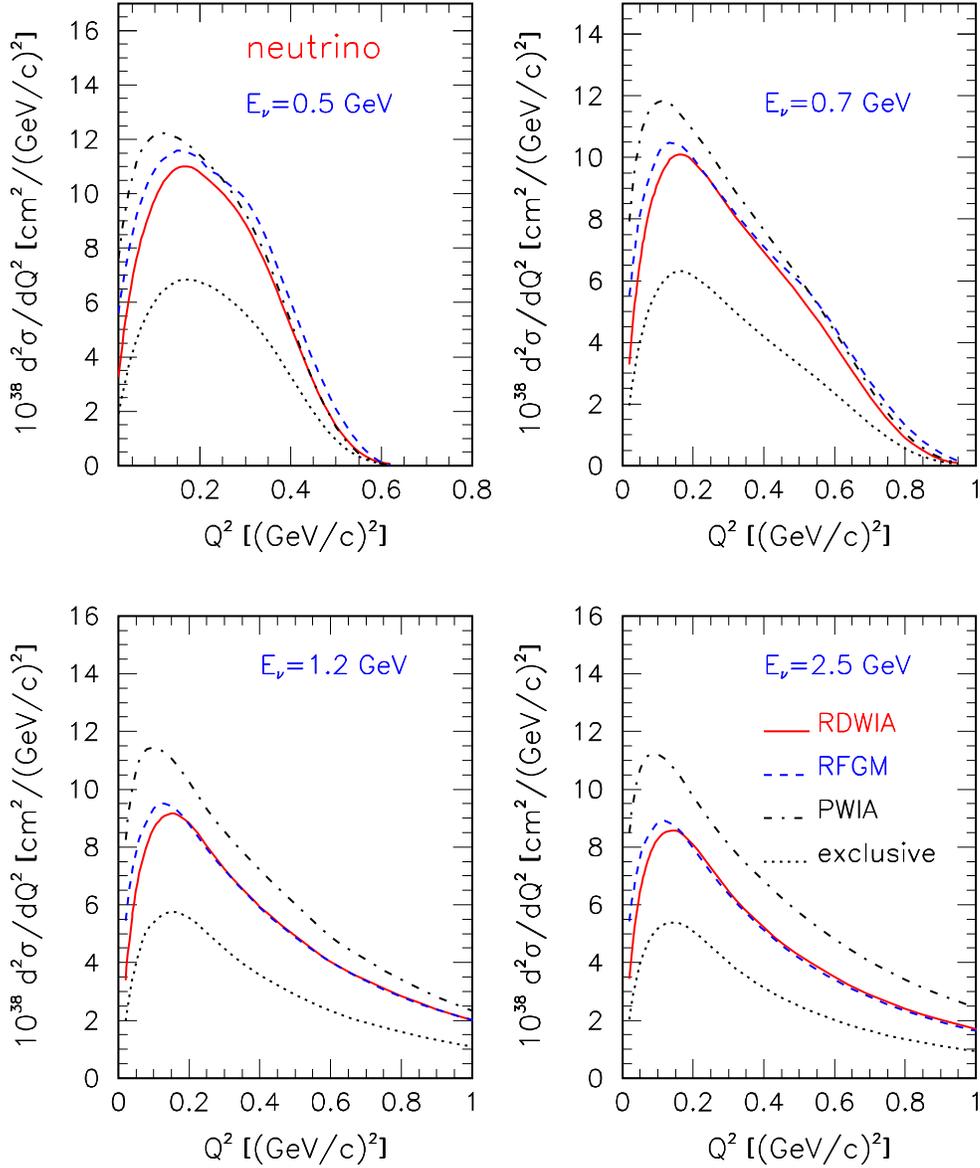}
  \end{center}
  \caption{(Color online) Inclusive cross section vs the four-momentum
    transfer $Q^2$ for neutrino scattering off ${}^{12}$C and for the four 
values of incoming neutrino energy:
$\varepsilon_{\nu}$=0.5,0.7,1.2, and 2.5 GeV. The 
solid line is the RDWIA calculation, whereas the dashed and dash-dotted lines 
are, respectively, the RFGM and PWIA calculations. The dotted lines are the 
cross sections for the exclusive reaction.}
\end{figure*}
\begin{figure*}
  \begin{center}
    \includegraphics[height=16cm,width=16cm]{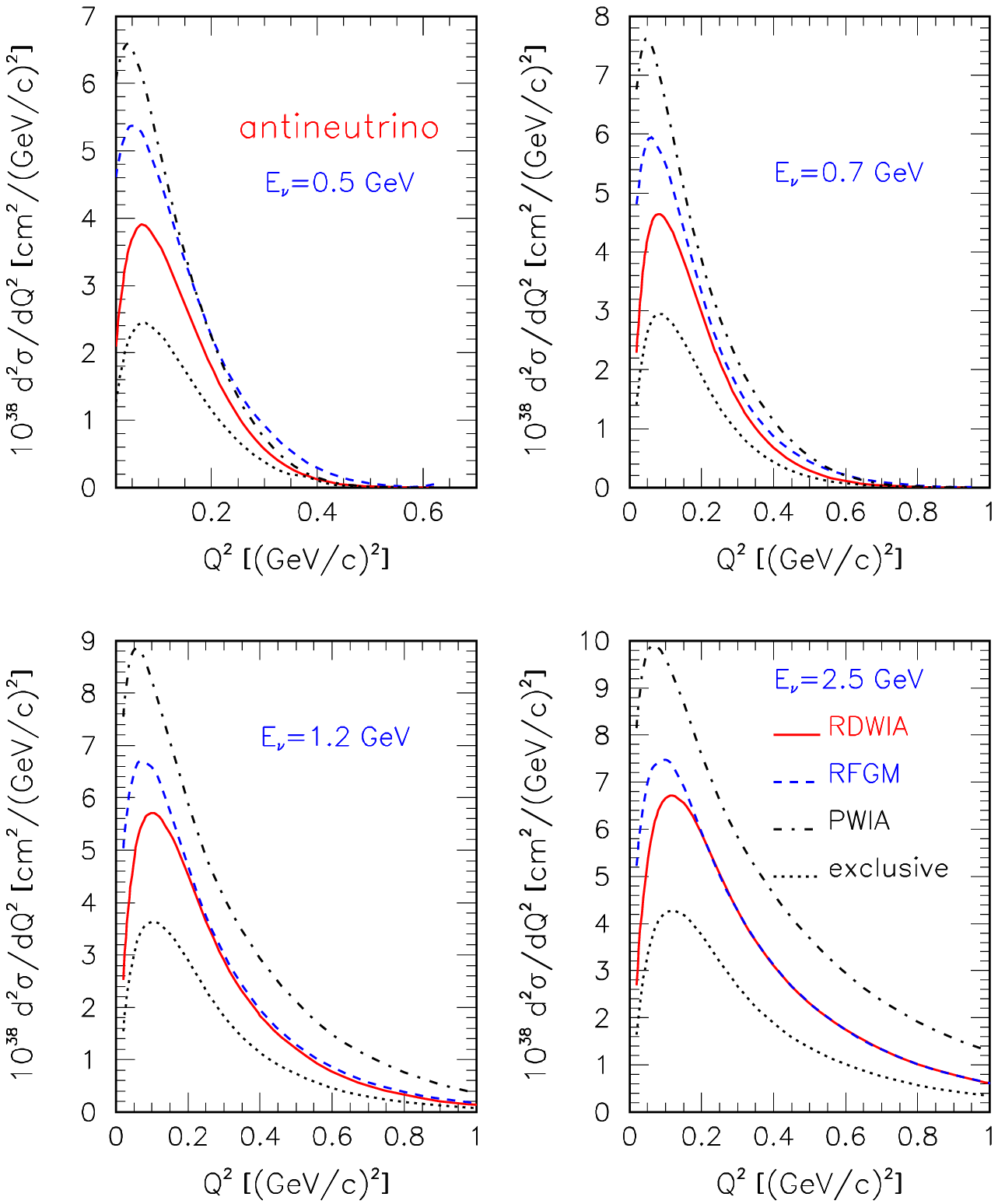}
  \end{center}
  \caption{(Color online) Same as Fig.4, but for antineutrino scattering.} 
\end{figure*}
\begin{figure*}
  \begin{center}
    \includegraphics[height=16cm,width=16cm]{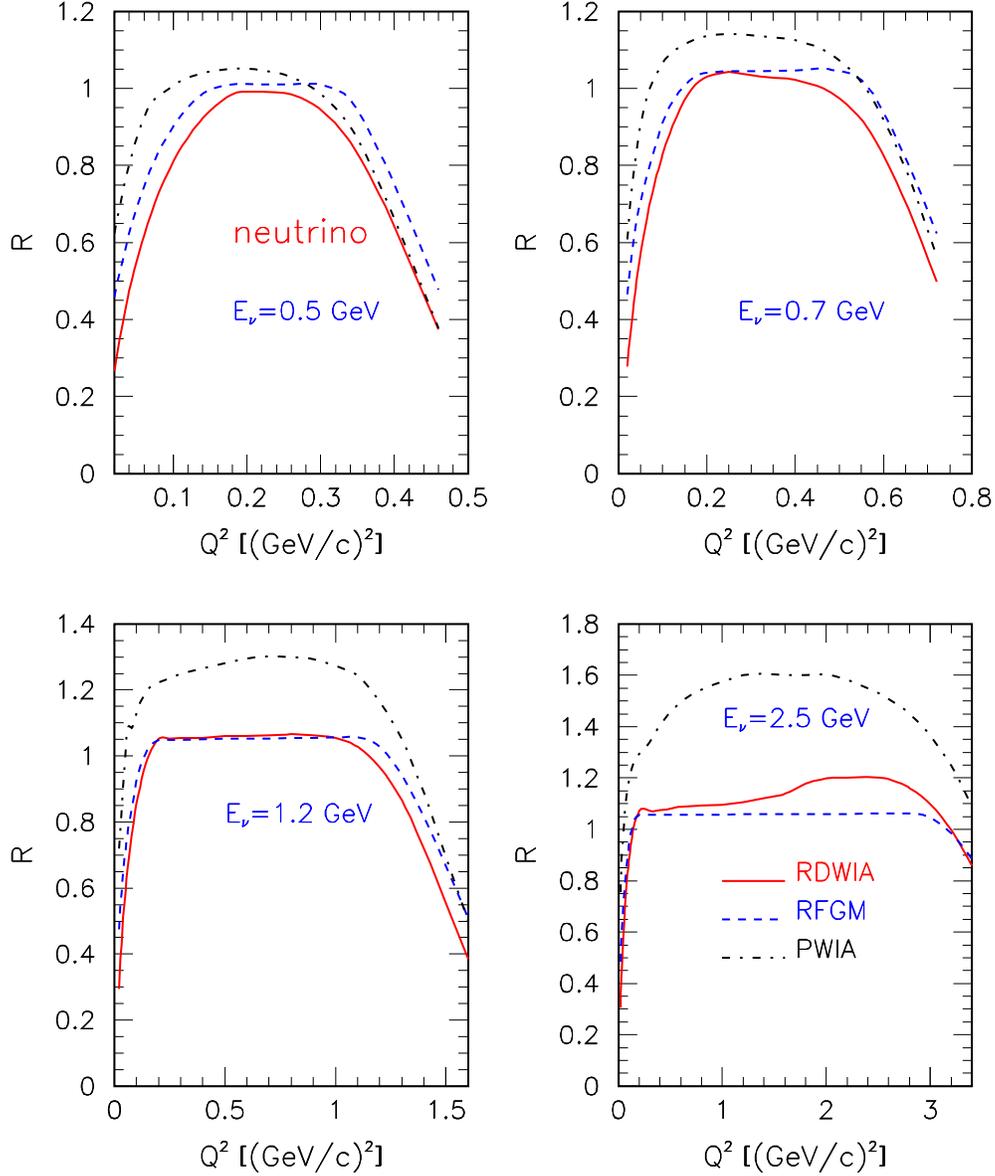}
  \end{center}
  \caption{(Color online) Ratio $R(\varepsilon_{\nu},Q^2)$ vs the four-momentum
    transfer $Q^2$ for neutrino scattering off ${}^{12}$C and for the four 
values of incoming neutrino energy: 
$\varepsilon_{\nu}$=0.5,0.7,1.2, and 2.5 GeV. As shown in the key, the cross 
sections were calculated with the RDWIA, PWIA, and RFGM.} 
\end{figure*}
\begin{figure*}
  \begin{center}
    \includegraphics[height=16cm,width=16cm]{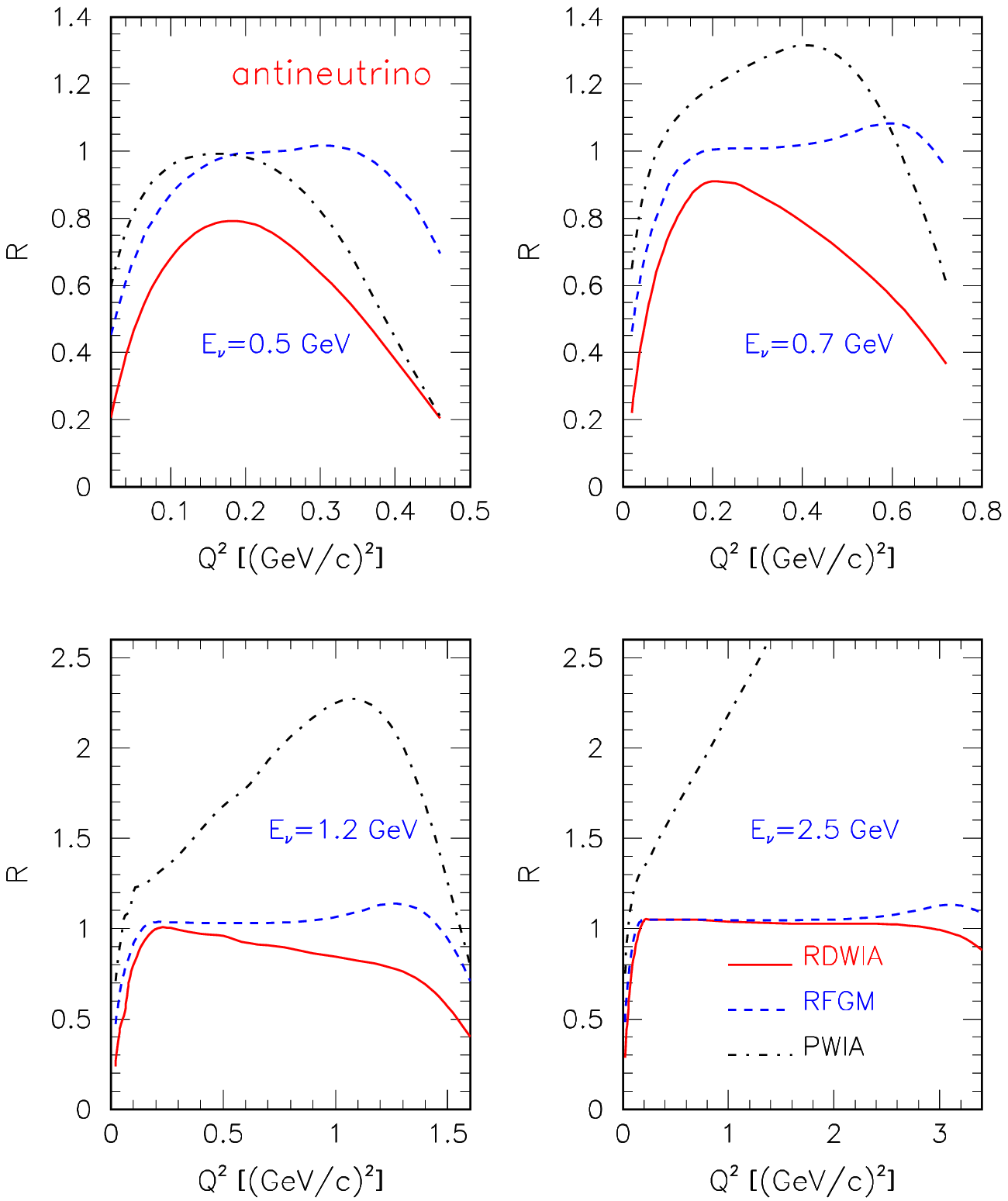}
  \end{center}
  \caption{(Color online) Same as Fig.7, but for antineutrino scattering.} 
\end{figure*}
\begin{figure*}
  \begin{center}
    \includegraphics[height=16cm,width=16cm]{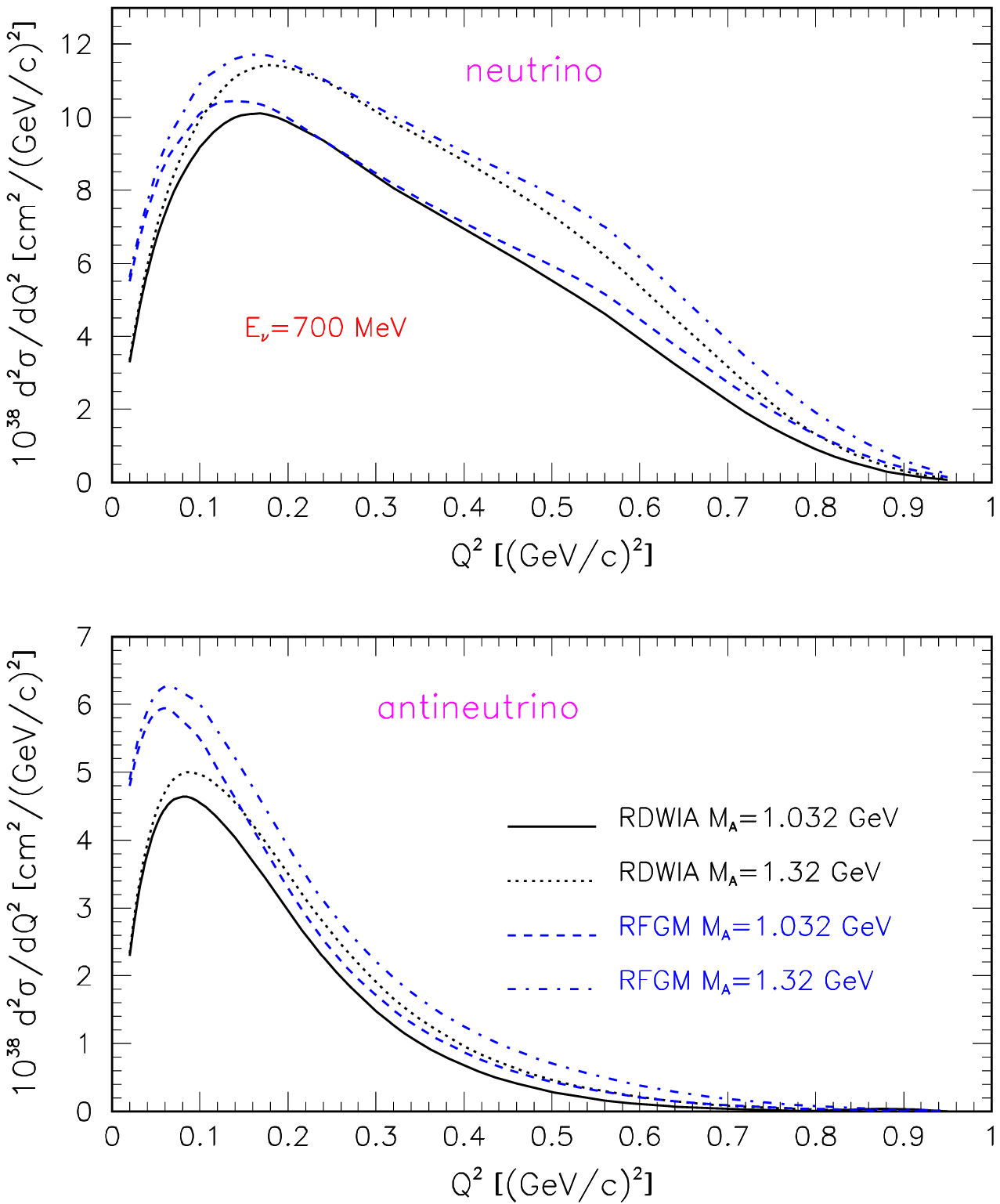}
  \end{center}
  \caption{(Color online) Inclusive cross section vs the four-momentum
    transfer $Q^2$ for neutrino (upper panel) and antineutrino (lower panel)
    scattering off ${}^{12}$C with energy $\varepsilon_{\nu}$=0.7 GeV and for
    the two values of axial mass $M_A$=1.032 and 1.32 GeV. As shown in the
    key, cross sections were calculated within the RDWIA and Fermi gas model.}  
\end{figure*}

To study nuclear effects on the $Q^2$-distribution, we calculated with
$M_A$=1.032 GeV the inclusive cross sections $d\sigma/dQ^2$ for
neutrino energies $\varepsilon_{\nu}$=0.5, 0.7, 1.2 and 2.5 GeV and compared
them with those for neutrino scattering on a free nucleon. The results for
neutrino and antineutrino scattering on carbon are presented in Figs. 4 and 5,
respectively, which show $d\sigma/dQ^2$ as functions of $Q^2$. Here, the 
results obtained in the RDWIA, are compared with cross sections calculated in 
the PWIA and RFGM. The cross sections for the exclusive reaction are shown as 
well. In the region $Q^2<$ 0.2 (GeV/c)$^2$ the Fermi gas model results for 
neutrino (antineutrino) are higher than those obtained within the RDWIA. At 
$Q^2$=0.1 (GeV/c)$^2$ this discrepancy equals 12\% (28\%) for 
$\varepsilon_{\nu}$=0.5 GeV and decreases to 7\% (12\%) for 
$\varepsilon_{\nu}$=2.5 GeV. The contribution of ($\nu,\mu N$) channels to the 
inclusive cross sections is about 60\%. 

Nuclear effects on the shape of the four-momentum transfer $Q^2$ distribution, 
i.e. ratio
$
R(\varepsilon_{\nu},Q^2)=(d\sigma/dQ^2)_{nuc}/(d\sigma/dQ^2)_{free},
$ 
where $(d\sigma/dQ^2)_{nuc}$ is the cross section scaled with number of 
neutron/proton in the target and 
$(d\sigma/dQ^2)_{free}$ is the cross section for (anti)neutrino scattering off
 free nucleon, are presented in Fig. 6 for neutrino and in Fig. 7 for 
antineutrino as functions of $Q^2$. Here, the results obtained in the RDWIA 
for energies $\varepsilon_{\nu}$=0.5, 0.7, 1.2 and 2.5 GeV are compared with 
those calculated in the PWIA and Fermi gas model. The nuclear effects are seen 
at low $Q^2$; the tail of the momentum distribution at high $Q^2$, an
overall suppression, and a slight change in the slope in the middle region at
$\varepsilon_{\nu} \geq$1 GeV can also be seen. The range of $Q^2$ where 
$R\approx const$, i.e. nuclear effects are negligible and, therefore cannot 
modify the value of $M_A$, increases with incoming neutrino energy. At energy 
higher than 1 GeV the range 0.3 $\leq Q^2\leq$ 1 (GeV/c)$^2$ can be used for 
$M_A$ extraction from $Q^2$ shape-only fit.           

We calculated $d\sigma/dQ^2$ cross sections at energy 700 MeV in the RDWIA 
and Fermi gas model with $M_A$=1.032 and 1.32 GeV. 
The results are shown  in Fig. 8 as functions of $Q^2$. Apparently  
at low $Q^2\leq$0.1 (GeV/c)$^2$ the cross sections depend weakly on the value
of the axial mass and $Q^2$ distributions are controlled by nuclear effects.   
\begin{figure*}
  \begin{center}
    \includegraphics[height=16cm,width=16cm]{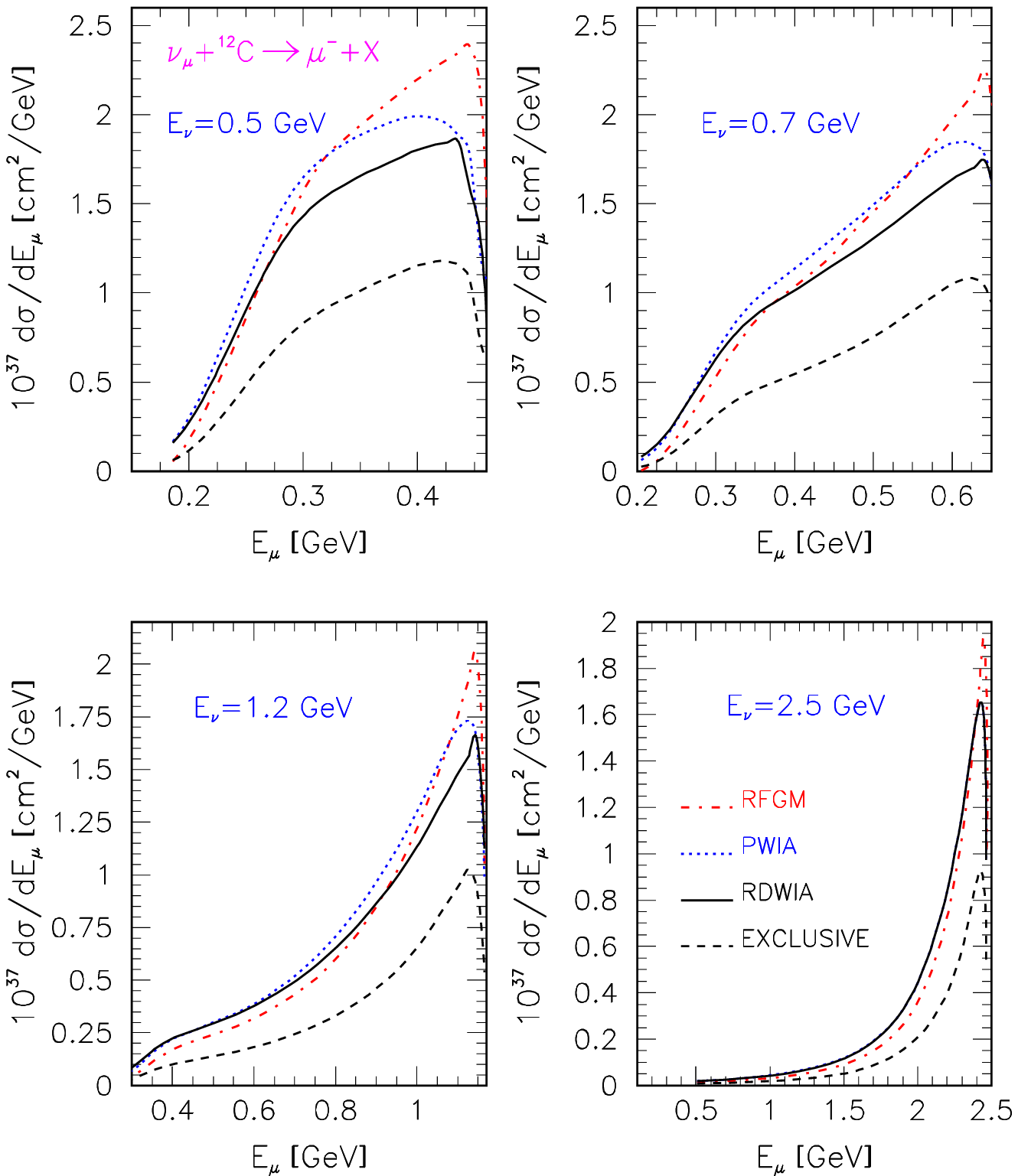}
  \end{center}
  \caption{(Color online) Inclusive cross section vs the muon energy for 
neutrino scattering on ${}^{12}$C and for the four values of incoming neutrino 
energy: $\varepsilon_{\nu}$=0.5,0.7,1.2, and 2.5 GeV. As 
shown in the key, the cross sections were calculated with the RDWIA, PWIA, 
RFGM, and RDWIA for the exclusive reaction.} 
\end{figure*}
\begin{figure*}
  \begin{center}
    \includegraphics[height=16cm,width=16cm]{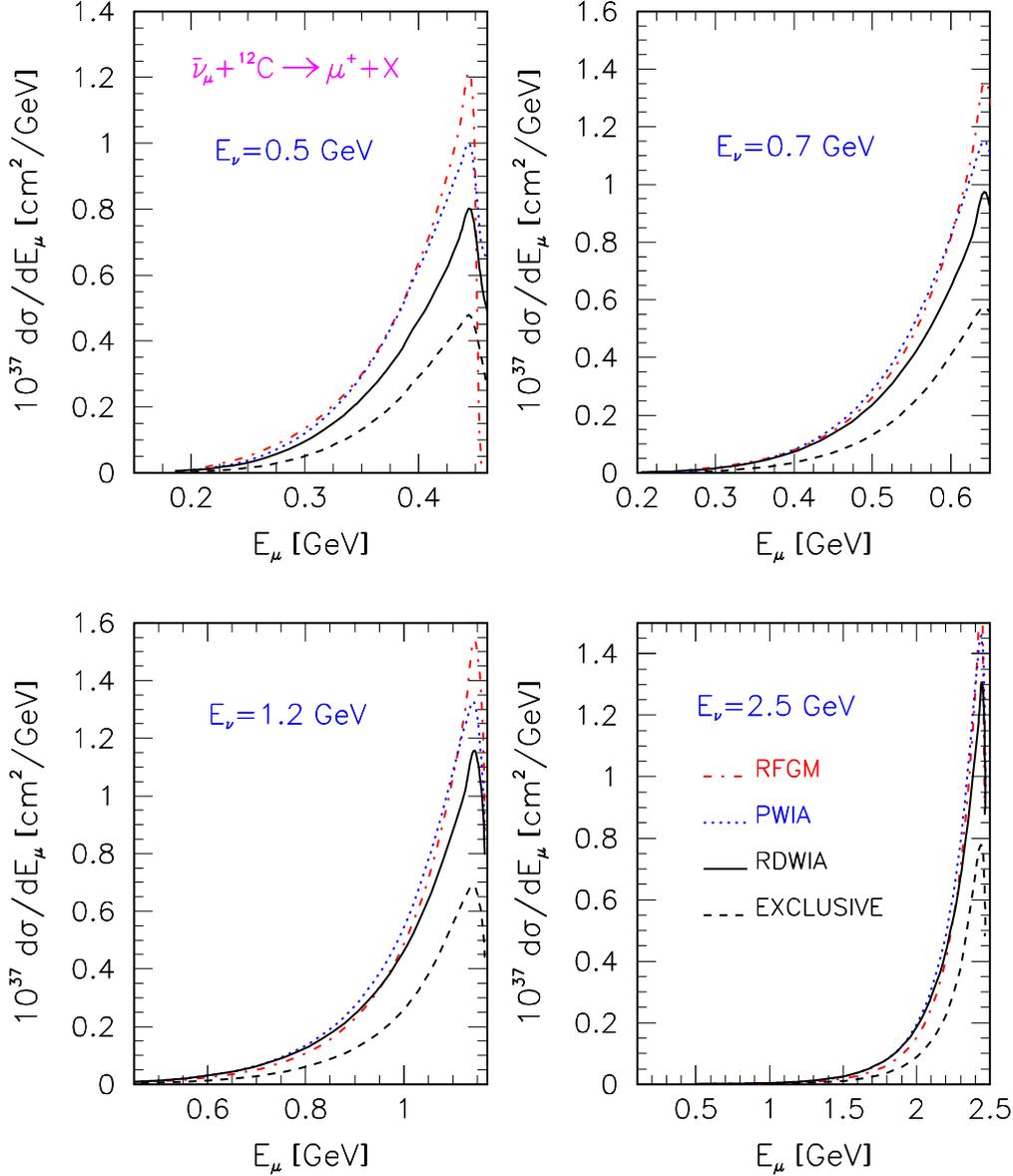}
  \end{center}
  \caption{(Color online) Same as Fig.9, but for antineutrino.} 
\end{figure*}
\begin{figure*}
  \begin{center}
    \includegraphics[height=16cm,width=16cm]{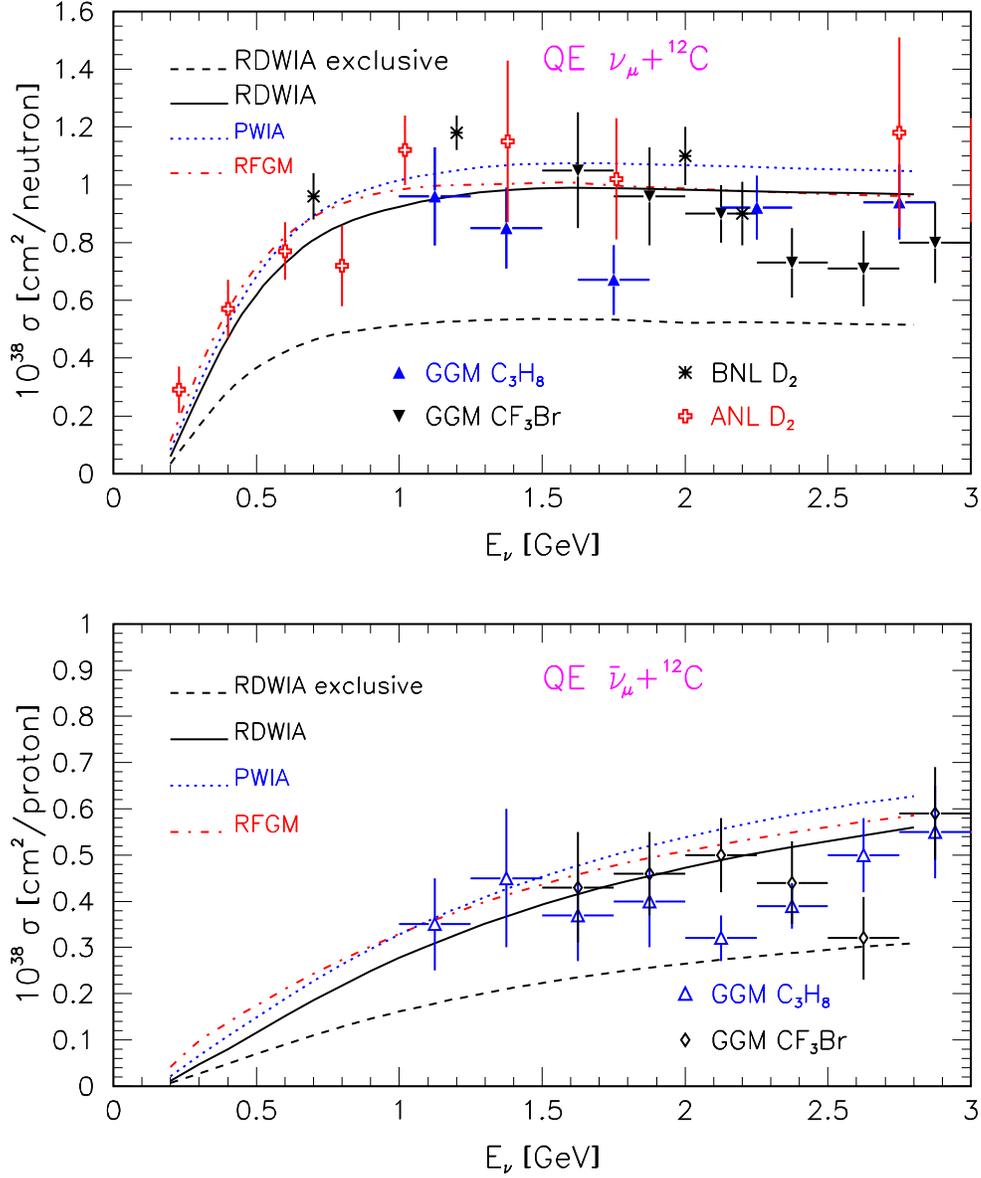}
  \end{center}
  \caption{(Color online) Total cross section for CC QE scattering of muon
neutrino (upper panel) and antineutrino (lower panel) on $^{12}$C as a
function of incoming (anti)neutrino energy. The solid line is the RDWIA 
result while the dashed-dotted and dotted lines are, respectively, the RFGM 
and PWIA results. The dashed line is the RDWIA result for exclusive reaction. 
Data points for different targets are from Refs.\cite{Mann, Baker, Pohl, 
Brunn}}. 
\end{figure*}
\begin{figure*}
  \begin{center}
    \includegraphics[height=14cm,width=14cm]{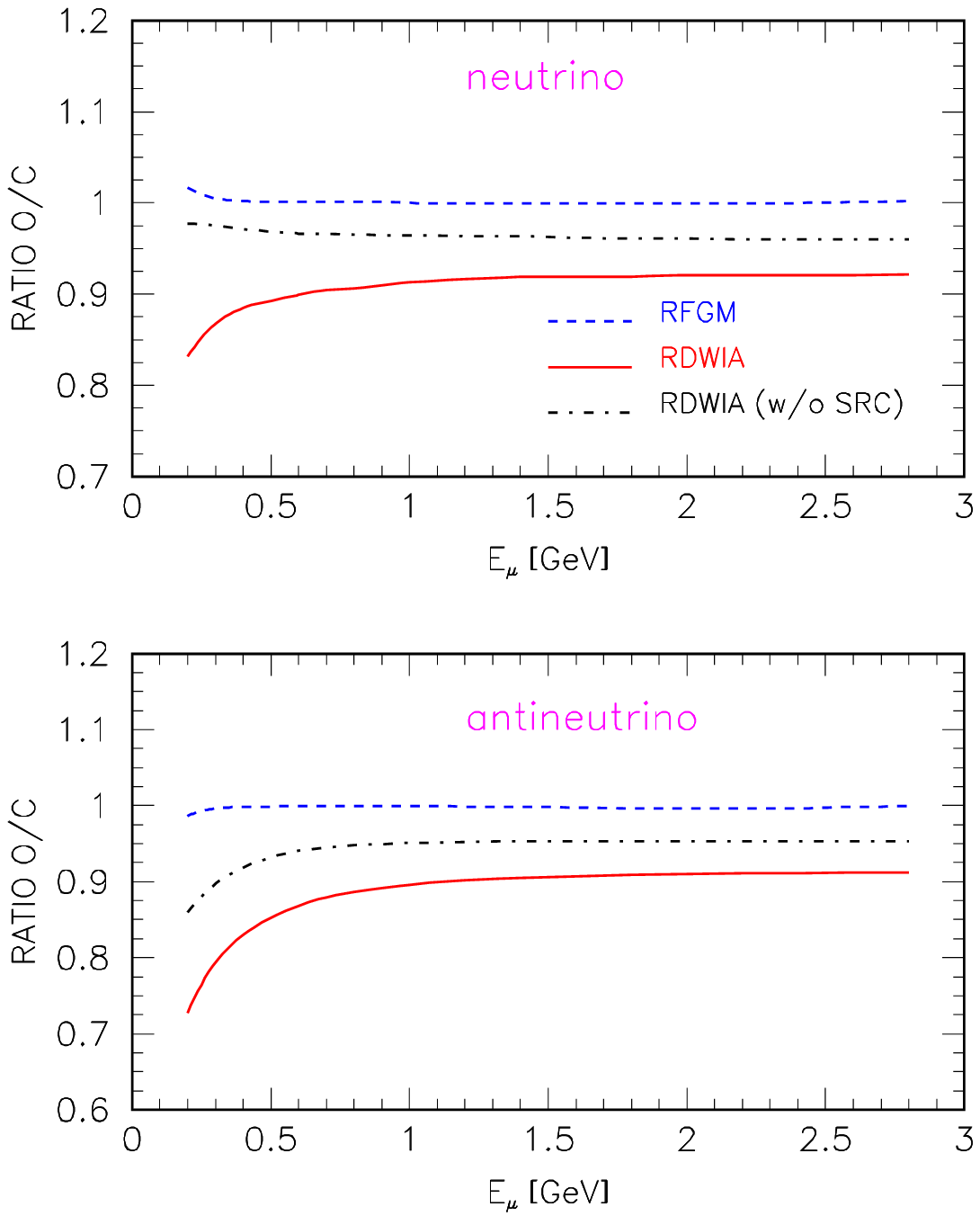}
  \end{center}
  \caption{(Color online) Ratio of the total cross sections per neutron/proton
 R=O/C for CC QE scattering of muon neutrino (upper panel) and antineutrino
    (lower panel) scattering on ${}^{16}$O and ${}^{12}$C vs incoming 
(anti)neutrino energy. The solid line is the RDWIA result, while the dashed and
    dashed-dotted lines are, respectively, the RFGM and RDWIA without 
contributions of the short-range correlations.} 
\end{figure*}

The inclusive neutrino and antineutrino cross sections for energies
$\varepsilon_{\nu}$=0.5, 0.7, 1.2, and 2.5 GeV are presented in Figs. 9 
and 10, which show $d\sigma/d\varepsilon_{\mu}$ as a function of muon energy. 
Here, the results obtained in the RDWIA with $M_A$=1.032 GeV are compared with 
the inclusive cross sections calculated in the PWIA, RFGM, and RDWIA for the 
exclusive reaction. The cross section values obtained in the RFGM
are higher than those obtained within the RDWIA. For neutrino
(antineutrino) cross sections in the region close to the maximum this
discrepancy is about 25\%(49\%) for $\varepsilon_{\nu}$=0.5 GeV and 
23\%(29\%) for $\varepsilon_{\nu}$=2.5 GeV. The contribution of $(\nu,\mu N)$ 
channels to the inclusive cross sections is about 60\%.

The neutrino and antineutrino total cross sections calculated with $M_A$=1.032
GeV up to neutrino energy 2.8 GeV, are shown in Fig. 11 together with data of 
Refs.\cite{Mann, Baker, Pohl, Brunn}. Also shown are the results obtained in 
the RFGM, PWIA as well as the contribution of the exclusive channels to the 
total cross sections. The cross sections are scaled with the number of 
neutron/proton in the target. The ratio between the neutrino cross sections 
calculated in the RFGM and RDWIA decreases with neutrino energy from about 
1.15 for $\varepsilon_{\nu}$=0.5 GeV to 1.02 for 
$\varepsilon_{\nu}$=2.6 GeV. For the antineutrino cross sections this ratio is 
about 1.5 for $\varepsilon_{\nu}$=0.5 GeV, and 1.05 for 
$\varepsilon_{\nu}$=2.6 GeV. The contribution of the exclusive channels is 
about 60\%. The results presented in Fig.11 show significant nuclear-model 
dependence for energy less than 1 GeV.

The RDWIA prediction for the CC QE flux-averaged total cross section is
compared with the experimental result from the LSND Collaboration at the Los 
Alamos for ${}^{12}$C$(\nu_{\mu},\mu^-)$ reaction~\cite{Auerbach}. The mean 
energy of the neutrino flux above threshold is 156 MeV. The calculated value of
10.14$\times$10$^{-40}$ cm$^2$ well agree with measured value of
(10.46$\pm$0.3$\pm$1.8)$\times$10$^{-40}$ cm$^2$. 

To compare the CC QE total cross sections for (anti)neutrino scattering on the
oxygen~\cite{But2} and carbon targets, we calculated the ratio 
$R(\varepsilon_{\nu})=(\sigma^O_{tot})_{nucl}/(\sigma^C_{tot})_{nucl}$, 
where the cross sections $(\sigma^i_{tot})_{nucl}$ are scaled with the number
of neutron/proton in the targets. The results obtained in the RFGM and RDWIA 
are shown in Fig. 12. The Fermi gas model predicts almost identical values of
$(\sigma^i_{tot})_{nucl}$ for ${}^{16}$O and ${}^{12}$C. In the RDWIA approach
the cross section calculated for oxygen is lower than that for carbon. For the 
neutrino (antineutrino) scattering this ratio is 0.90(0.88) at 
$\varepsilon_{\nu}$=0.7 GeV and 0.92(0.91) at $\varepsilon_{\nu}$=2.6 GeV.

To study the $NN$-correlation effects, we calculated the ratio
$R(\varepsilon_{\nu})$ without the $NN$-correlation contribution, i.e. with
$S_{\alpha}$=1 for all bound nucleon states in the oxygen and carbon
targets. The difference between results obtained with and without the
high-momentum component contribution is about 5\% for $\varepsilon_{\nu}\geq$1 
GeV. In Ref.\cite{But2} was shown that the $NN$-correlation effect
reduces the total cross section in proportion to the missing strength in the
nuclear ground state, which is about 25\% for ${}^{16}$O and 11\% for
${}^{12}$C. 

Therefore in the long-base line neutrino oscillation experiments a part of the
near detector must include some of the same target material, as the far
detector to reduce the systematic uncertainty due to nuclear effects on the CC
QE total cross section.

\section{Conclusions}

In this paper, we study electron and CC quasi-elastic (anti)neutrino
scattering on a carbon target in different approximations (PWIA, RDWIA,
RFGM) placing particular emphasis on the nuclear-model dependence of the
results.
In RDWIA, the LEA program, adapted to neutrino interactions, was used to
calculate the differential and reduced exclusive cross sections.
We found that the reduced cross sections for (anti)neutrino scattering are
similar to those of electron scattering and the latter are in a good
agreement with electron data.
In calculating the inclusive and total cross sections, the imaginary
part of a relativistic optical potential was neglected and 
the SRC effect in the target ground state was taken into account. This
approach was tested against electron-carbon inclusive scattering data. This  
test revealed an overall agreement with the data, with the differences between
calculated and measured cross sections in the peak region less than 12\%.

We calculated $d\sigma/dQ^2$ cross sections for different neutrino energies
and estimated the range of $Q^2$ where nuclear effects on the shape of $Q^2$
distribution are negligible. Also was shown that at low $Q^2<$0.1 (GeV/c)$^2$ 
the coss sections depend weakly of the values of the axial mass.   

The CC QE total cross sections predicted by the RFGM are higher than the
corresponding values obtained in the RDWIA and this difference decreases with
neutrino energy. The flux-averaged total cross section was calculated within
the RDWIA approach and compared with the experimental result from the LSND
Collaboration. The calculated cross section is in good agreement with
data. We compared the CC QE total cross sections (scaled with the number of
neutron/proton in the target) for (anti)neutrino scattering on the oxygen and 
carbon targets and found that the cross sections calculated within the RDWIA 
for oxygen, are lower than those calculated for carbon, and the SRC effects 
increase this difference.  

We conclude that the data favor the RDWIA results. 
This indicates that the use of RDWIA in Monte Carlo simulations
of the neutrino detector response would allow one to reduce the systematic
uncertainty in neutrino oscillation parameters.

\section*{Acknowledgments}

The author greatly acknowledges S. Kulagin, J. Morfin, G. Zeller,
N. Jachowicz, M. Wascko, R. Gran, and T.Katori for fruitful discussions at 
different stages of this work.  
%


\end{document}